\documentclass[10pt,conference]{IEEEtran}
\usepackage{amsmath,amsfonts}
\usepackage{algorithmic}
\usepackage{algorithm}
\usepackage{array}
\usepackage[caption=false,font=normalsize,labelfont=sf,textfont=sf]{subfig}
\usepackage{textcomp}
\usepackage{stfloats}
\usepackage{url}
\usepackage{verbatim}
\usepackage{graphicx}
\usepackage{cite}
\begin{document}

\title{The Mythical Good Software}

\author{\IEEEauthorblockN{Aydin Homay}
\IEEEauthorblockA{\textit{Chair of Industrial Communications} \\
\textit{Technische Universität Dresden}\\
Dresden, Germany \\
https://orcid.org/0000-0002-6425-7468}
}

\markboth{Journal of \LaTeX\ Class Files,~Vol.~14, No.~8, August~2021}%
{Shell \MakeLowercase{\textit{et al.}}: A Sample Article Using IEEEtran.cls for IEEE Journals}


\maketitle

\begin{abstract}
Good software has high cohesion and low coupling is clumsy, obscure, and in some certain cases could be actually a harmful state of being. It is clumsy because there is no perfect correlation between higher cohesiveness and optimum design, and it is obscure because it conveys the message that coupling and cohesion are two distinct design principles, while there are in principle the same design approaches, and only the time and space differ between them, and it could also be a harmful state of being because we should not always aim for higher cohesiveness without considering its cost. 

In the course of this study, we aim to elucidate for the readers the meaning and underlying philosophy of the aforementioned paragraph.  
\end{abstract}


\section{Introduction}
\IEEEPARstart{S}{oftware} systems are the source of chaos!~\cite{Booch-2007}, as are the business systems. Although different battles began between order and chaos, the war never ends. Order in a system is so exciting because it facilitates moderation (e.g., changeability, predictability) of systems, especially when they tend to be complex. However, it is not a free lunch as it comes at its own cost, and, of course, the cost differs from system to system.

The larger the system size, the higher the cost of introducing the order. As systems grow (incorporating more features and targeting a larger user base), they become more difficult to modify. It should be noted that even the original designers may struggle to grasp larger systems~\cite{Miller-1956}.

Is size the only concern for the designer? The answer is no. A software system might be small, yet very expensive and challenging to modify, while a larger system might continually evolve at a lower cost. Thus, there are additional factors contributing to complexity that need to be managed. 

In 1965 E. W. Dijkstra~\cite{Dijkstra-1965}, discussed using the concept of "divide et impera" (divide and rule) to master the complexity of the software. The earlier approach he suggested was to divide the program based on its algorithmic flow chart.  

It was 1971, when David L. Parnas argued that dividing code based on a flow chart (which Dijkstra introduced) is not efficient. He introduced the idea of breaking software down into modules in the direction of changeability~\cite{Parnas-1972}.

In 1978~\cite{Yourdon-1978}, Yourdun and Constantine suggested that prior to any decomposition, its cost should be assessed, as the expenses associated with decomposition (dividing) could exceed the benefits of avoiding it. 

Today, this technique in the software field is referred to as modularization (componentization) and has been widely used to address some complexities in software systems. However, the ideal size of each module is still a topic of debate.~\cite{Broy-2018}. Perhaps the latest debate was raised by Microservices~\cite{Newman-2015} in 2015.

Although it is commonly accepted that every module (i.e., component, class, or service) should maintain strong cohesiveness and minimal coupling with other modules, the definitions of "high" and "low" are ambiguous. It remains unclear just how high it is considered sufficient and how low it is deemed adequate.

The main argument of this paper is to first clear the common misunderstanding about modularization (granularity vs. interchangeability), then explain the ambiguity behind the "high cohesion and low coupling" principle, and finally illustrate why cohesion could eventually be a harmful state of being.

The subsequent sections of this paper are arranged as follows: Section II gives an overview of modularization. Section III delves into Coupling and Cohesion, while Section IV explores the "low coupling and high cohesion" principle. Section V addresses the drawbacks of cohesion by explaining the fundamental theory of software engineering. Section VI presents the main argument of this study based on five corollaries. And Section VII concludes our argument.

\section{Journey of Modularization}
\label{Modularization}\
The topic of modularization started in housing (building construction) in 1932~\cite{Russel-2012}, then found its way into electronic and manufacturing during 1940-1950~\cite{Baldwin-Clark-2000, Russel-2012}, and eventually into software development from 1960~\cite{Baldwin-Clark-2000, Russel-2012}.

In late 1957, a group of IBM engineers in Endicott, New York, created the foundations of what they called the Standard Modular System (SMS). SMS modules were built to consistent specifications, with standardized dimensions and connectors, allowing IBM to produce "pluggable" circuit boards that could be easily swapped out or replaced without affecting the overall system~\cite{Baldwin-Clark-2000}. 

It is important to note that starting 1950 at IBM research and engineering work was focus around hardware and the emphasis was not on reusability because computer hardware was not yet being produced on a large scale, instead, the priority was on the possibility of changing electronic parts, specifically replacing a defective part with a new one without effecting the entire circuit. The result of this was the successful product IBM’s System/360~\cite{Baldwin-Clark-2000} which unlike the software part (IBM System/360 Operating System), was a great success~\cite{Brooks-1972}.

\subsection{Routine, Subroutine, Macro, and Plex}
In 1950 when Tompkins et al. attempted to manage the complexity of problems by dividing them into smaller sub-problems in order to reduce the need for computational resources. The limitations arising from constraints of computing machines led Tompkins to attempt to divide the problems into smaller separate subproblems~\cite{Tompkins-1950}.

In 1952, Hopper tried to take a similar approach by segmenting and encapsulating codes into smaller sections (macros) with specific functionality to enhance \textbf{reusability}, accuracy, and reduce complexity (in terms of execution time) in the first operational "compiler" named A-0. In her paper~\cite{Hopper-1952}, Hopper referred to this idea as "compiling routines". 

However, the specific term "Macro" came into more attention when Barnett explored this subject in his paper "Macro-Directive Approach to High-Speed Computing" in 1959~\cite{Barnett-1959}. 

Further advancements in the concept were detailed in 1960 through other work such as "Macro instruction extensions of compiler languages"~\cite{McIlroy-1960}, where McIlroy demonstrated macro with the following example:
\begin{verbatim}
    ADD, A, B, C = FETCH, A
                   ADD, B
                   STORE, C
\end{verbatim}
ADD is the name of the macro, and A, B, and C are arguments to be provided during the invocation of ADD.

In 1961, D. T. Ross introduced the concept of a "Plex-Structure," contrasting it with list and tree data structures. He claimed that although lists and trees can be advantageous in some situations, they can also induce substantial complexity, particularly when it comes to processing symbolic and numerical data at the same time. Ross asserted that a "Plex-structure" could address these inefficiencies by consolidating all interrelated data into a single, unified component which he called it the "Plex"~\cite{Ross-1961}.\footnote{The authors believe that this work had a significant impact on the development of graph and object storage techniques.}

The idea of reducing the size of program in order to manage its complexity (in terms of execution and development) started to gain more attention by mid 1960 when Dijkstra wrote "Programming considered as a human activity" in 1965~\cite{Dijkstra-1965} and he discussed about "divide-and-rule" approach as a method of dealing with program complexity. However, it is important to note that he did not employ the words "modular" or "modularization" to express the idea that his concept of "divide-and-conquer" effectively refers to modularization. Indeed, Dijkstra perceived modularization as a method for structuring knowledge~\cite{Dijkstra-1974}\footnote{Parnas shared a similar vision through the concept of information encapsulation~\cite{Parnas-1972}}.

\subsection{Modules and Components}
Influenced by modularization in hardware systems\cite{Gray-1957, Baldwin-Clark-2000}, in software systems also modules were smaller than programs and theoretically a program could contain multiple modules, where each module might incorporate several routines and each routine could include multiple subroutines\cite{Baldwin-Clark-2000}.

Larry L. Constantine authored a paper in the March 1967 issue of "Computer and Automation"~\cite{computer-and-automation-1967-march}, titled "A Modular Approach to Optimization"~\cite{Constantine-1967}. In this paper, he evaluated how modularization could improve both the efficiency of program execution and the development process, such as the time needed for debugging. However, he did not detail a structured methodology for implementing modularization. Note that this is the beginning of thinking about modularization in software with the intention of mastering its complexity in terms of execution time, power of resource, and development and troubleshooting effort.

Unlike Constantine, McIlroy, Brown, and Hopkins tried to exhibit modularization by revolving it around making software more \textbf{changeable}, adaptable, and portable by an idea introduced as in their paper called "Mass Produced Software Components"~\cite{NATO-1968} presented during NATO conference in 1968\cite{NATO-1968}, and continued in NATO conference 1969 by W.S. Brown, extending it to software portability~\cite{NATO-1969}.

McIlroy included the attributes of "modules" within the definition of "components" and added additional characteristics such as "precision, robustness, generality, time-space behavior, algorithm, and interface." He emphasized that these components should be compilable by end-users to ensure that they operate correctly on the user hardware (portability)~\cite{NATO-1968}\footnote{Readers should note that NATO-1968 conference occurred during a time when it was called as software crises e.g., the IBM OS/360 faced failure, incurring significantly greater costs for IBM compared to the IBM 360 hardware~\cite{Brooks-1986}.}.

The characteristics McIlroy identified for defining software components in 1968~\cite{NATO-1968} later significantly influenced the criteria for quality software (reliability, maintainability, changeability, generality, and efficiency) discussed greatly by Myers in his 1975 book "Reliable software through composite design"~\cite{Myers-1975}, which gained significant acceptance.

Although his original work around Plex was heavily related to data structures, D. T. Ross introduced a new vision of Plex that went beyond the data structure and consists of three parts: Data, Structure, and Algorithm (that is, behavior) in 1968 during NATO conference\cite{NATO-1968}. He argued that we need all three aspects to build a complete model in programming world that potentially could represent an element in the real world, and to do so it is not sufficient to just talk about data structures, though this is often what people do. The structure shows the interrelationships of pieces of data, and the algorithm shows how these structured data should be interpreted. For example, a data structure could stand for two different things if you didn’t know how to interpret it. 

The key thing about the plex concept is that it tries to capture the totality of meaning, or understanding, of some problem of concern in some way that will map into different mechanical forms, not only on different hardware, but also using different software implementations.

D. T. Ross achieved a significant milestone with Plex in his paper from 1976~\cite{Ross-1976} where he developed Plex towards the Object-Oriented Language paradigm by binding data and procedures together.

In 1971, David L. Parnas authored an influential essay called "On the Criteria To Be Used in Decomposing Systems into Modules"~\cite{Parnas-1972}. This paper discusses and compares two methods for examining and organizing the logic behind a basic algorithm. The conclusion of the paper presents the following.

"We have attempted to show with these examples that starting a system's decomposition into modules with a flowchart is almost always erroneous. Instead, we suggest starting with a list of challenging design decisions or those likely to change. Each module is then crafted to conceal this decision from the others." In fact, Parnas concludes that modules should be divided according to their potential changes, and by examining these changes, the distinctions between modules can be understood."

Parnas's significant contribution was integrating business requirements into the program design process and accounting for potential changes in program code modularization, instead of depending on the previously used flowchart-oriented modularization, as discussed by Dijkstra.

In 1974, E. W. Dijkstra wrote another classic paper in which he coined the "separation of concerns" and the "single responsibility"~\cite{Dijkstra-1974} to be considered for addressing problems in computing science more systematically and intelligently. 


By 1989, when Chris Reade authored his book entitled "Elements of Functional Programming"~\cite{Reade-1989}, which elaborates on the idea of separation of concerns, it was clear that the concept of separation of concerns coined by Dijkstra was gaining acceptance.

\subsection{Ultimate Goal of Modularization}
Influenced by modularization in hardware, the main objective of modularization in software, as well as subjects such as macros, routing, and subroutines, in its early adaptation was focused on managing complexity primarily in the context of execution time, memory usage, computing response, and development processes (coding and debugging) and later on reusability. 

However, after the late 1960s, notably from the NATO conferences of 1968 and 1969~\cite{NATO-1968,NATO-1969}, the focus on the objectives of modularization expanded beyond reusability, covering aspects of changeability and portability as well.

The core philosophy behind modularization from 1970 is not to master the kind of complexity that Tompkins and others discussed in early 1950~\cite{Tompkins-1950,Hopper-1952, Barnett-1959,Dijkstra-1965}, but a type of complexity that has persisted and will continue to exist in a domain that reuseability cannot address (i.e. creating evolutionary software).

As an author's note, although modularization in software systems begins with decomposing, the designer can eventually utilize composition and oscillate between different levels of granularity, depending on the concerns guiding the system's design with the goal of reaching an acceptable balance among varying trade-offs. \footnote{This argument could be supported by addressing Chapter 13 of "Clean Architecture" authored by C. R. Martin~\cite{R.C.Martin-2017}}.

\subsection{Modularization Dilemma}
We will begin by examining two methods that have been used by researchers and professionals for modularization. We will refer to the first method as the Parnas method~\cite{Parnas-1972} focusing on information encapsulation and changeability, and the second method as the Constantine method focusing on increasing cohesion and reducing coupling.

Parnas' method involves breaking down software into modules according to specific criteria or concerns such as changeability. For instance, in his 1971 paper~\cite{Parnas-1972}, he states that:

"The decision to divide a system into n modules of a given size does not determine the decomposition, unless we define clear criteria to use for decomposition, such as rewriting the module with little knowledge about its internals and allowing the new module to be reassembled and replaced without reassembling the whole system."    

Constantine's method forms around the "Plex" concept introduced by D.T. Ross~\cite{NATO-1968}. For example, W. Stevens, G. Myers and L. Constantine published a paper in the IBM Systems Journal~\cite{Constantine-1974} in 1974 arguing that dividing software systems into smaller cohesive parts reduces software complexity. The following quote has been repeated from this publication. 

"Mr. Constantine has observed that the programs that were the easiest to implement and change were those composed of simple independent modules. The reason for this is that problem solving is faster and easier when the problem can be subdivided into pieces that can be considered separately.

A similar argument is observed from DeMarco's 1979 book~\cite{DeMarco-1979} based on Constantine's work on software modularization:
"Obviously, the larger the system, the more complex the analysis. There is little we can do to limit the size of a system; there are, however, intelligent and unintelligent ways to deal with size. An intelligent way to deal with size is to partition. That is exactly what designers do with a system that is too large to deal with conveniently so they break it down into component pieces (modules)."

In fact, Constantine, along with colleagues such as Yourdon, DeMarco, and Page-Jones, used the coupling and cohesion principle to guide modularization with the intention of reducing the complexity of software in terms of human ability to understand and troubleshoot software systems\footnote{They relied on average human  short-term memory which can hold in 7±2 elements~\cite{Miller-1956}}, reuseability to improve software development speed, and modifiability to improve software maintainability~\cite{Yourdon-1978, DeMarco-1979, Page-Jones-1980}.

On the other hand, Meyer~\cite{Meyer-1997b}, R. C. Martin~\cite{R.C.Martin-2000, R.C.Martin-2017} and G. Booch~\cite{Booch-2007} which do not adhere to Constantine's method in its original form aligns more closely with Parnas and Dijkstra's understanding about modularization which is based on concerns and information hiding (encapsulation). They do not explicitly endorse the coupling and cohesion spectrum nor rely merely on the "high-cohesion and low-cohesion" principle.

Although distinguishing between the works of Parnas and Constantine might initially seem trivial or even non-existent, the rest of the paper will demonstrate how these two ideas lead to different modularization.

It is important to note that the basis for software evolution management was laid in the 1980s on the basis of its changeability aspects by Lehman when he defined laws of software evolution and, among others, identified that systems are subject to dynamics, causing continuing changes resulting in increased complexity~\cite{Lehman-1980}.

Currently, numerous principles are widely discussed and accepted ~\cite{Yourdon-1978,Gamma-1994, Evans-2003, Fowler-2001, Pressman-2015, Lethbridge-2005, Newman-2015} to assist software engineers in optimizing modularization by determining the appropriate granularity and the right degree of functionality and responsibility for each module. However, most of them are biased to the "high cohesion and low coupling" principle as a heuristic method to find the right granularity. 

But the central question is whether this principle is accurate and robust enough to guide software development towards enhanced modularization. This issue will be explored further in the following section.

\section{Coupling and Cohesion}
When examining the definitions of 'coupling' and 'cohesion' ('strength' as described by Myers~\cite{Myers-1975}), it becomes apparent that both terms address the same concept: coupling software elements. The differences lie in the dimensions of space and time, specifically where, when, and how. However, most software engineers regard these as separate subjects and discuss them separately.

\textbf{Coupling} is defined as the degree of intermodular relatedness (that is, the relationship among modules)~\cite{Constantine-1974, Myers-1975,Yourdon-1978,DeMarco-1979, Page-Jones-1980,Pressman-2015, Kent-2023,Newman-2015,Evans-2003}. This means module A is coupled to module B if changing module A leads to change module B~\cite{Kent-2023}.

\textbf{Cohesion} is defined as the degree of intramodular relatedness (that is, the relationship among elements of the same module)~\cite{Constantine-1974, Myers-1975,Yourdon-1978,DeMarco-1979, Page-Jones-1980,Pressman-2015, Kent-2023,Newman-2015,Evans-2003}. This means that module A contains elements that are in close relation (tight coupling) with each other, and all of them are essential for the certain behavior that is expected from the existence of module.

\subsection{background}
In 1978, Yourdon and Constantine dedicated two chapters of their book "Structured Design"~\cite{Yourdon-1978} to "Coupling" and "Cohesion". Regrettably, it is a common misconception that the book "Structured Design" originated these terms and formed the "high cohesion and low coupling" design principle. Although Constantine contributed to this topic significantly~\cite{Constantine-1974} our review of the literature shows that these terms (coupling and cohesion) existed a decade earlier~\cite{NATO-1968} and even Myers in his 1975 book "Reliable software through composite design"~\cite{Myers-1975}, covered module coupling and module strength (cohesion) in separate chapters and introduced the idea of "low coupling and high strength".\footnote{Note: we fully acknowledge Larry L. Constantine's groundbreaking work on coupling and cohesion, our only intent is to point out that the book "Structured Design" was not the original source to introduce these concepts.} 

The primary distinction between "Reliable software through composite design"~\cite{Myers-1975} and "Structured Design"~\cite{Yourdon-1978} lies first in the terminology and slight variations in the degrees of coupling and strength (cohesion) introduced in each book, and second, in contrast to Yourdon and Constantine, Myers believes that there is a perfect correlation between module coupling and module strength, where an increase in one results in a decrease in the other.

Apparently, Constantine favored the term "cohesion" instead of "strength" because of his sociological background. It is important to note that he earned his diploma in "Family Therapy" from the Boston Family Institute and was actively publishing papers in this field as well.

In 1979, Tom DeMarco also discussed coupling and cohesion in his book "Structured Analysis and System Specification"~\cite{DeMarco-1979} and interestingly in terms of coupling he relied on \cite{Myers-1975} and in terms of cohesion on \cite{Yourdon-1978}. 

In 1980, M. Page-Jones~\cite{Page-Jones-1980} also devoted two chapters of his book "The Practical Guide to Structured Systems Design" to the topics of coupling and cohesion. However, the second edition~\cite{Page-Jones-1988} uses a nearly identical approach to the one discussed in "Structured Design", categorizing various types of couplings comparable to those shown in Table\ref{tab:classic_coupling}.

As observed, almost everyone who discusses coupling and cohesion in software products has treated them as separate subjects and analyzed them individually~\cite{Myers-1975, Yourdon-1978, DeMarco-1979, Page-Jones-1980, Lethbridge-2005, Pressman-2015}. 

We contend that this was a fundamental error, as it suggests that coupling and cohesion are distinct topics while, in reality, both pertain to the strategic use of coupling in software design, differing only in where coupling should be applied and where it should be avoided, where it should be accepted, where it should be rejected, where it should be maximized, where it should be kept to minimum.

\subsection{Coupling}
According to Britannica, coupling refers to the concept of machinery that connects two rotatable shafts. In the field of physics, the term 'coupling' describes the interaction between two entities. In classical mechanics, it signifies the connection between oscillating systems, such as pendulums linked by a spring. In particle physics, two particles are considered coupled if they interact through one of the four fundamental forces.

As we discussed earlier, the problem of coupling (as the cause of low maintainability and changeability) was observed well before the principle was established. Dijkstra wrote in 1968~\cite{Dijkstra-1968a} a classic paper against the "GoTo" statement, pointing out that the transfer of control from one module to another module has disastrous effect. Later, this was called "\textit{Control Coupling}"~\cite{Myers-1975, Yourdon-1978, DeMarco-1979,Page-Jones-1988}.

In software systems, Myers~\cite{Myers-1975}, Yourdon and Constantine ~\cite{Yourdon-1978}, DeMarco~\cite{DeMarco-1979}, and Page-Jones ~\cite{Page-Jones-1988} define coupling as the degree of intermodular relatedness (that is, the relationship among modules). They further differentiate between highly coupled, loosely coupled, and decoupled intermodular relatedness, indicating that highly coupled modules are strongly interconnected, loosely coupled modules have weak interconnections, and uncoupled or decoupled modules have no interconnections. 

The Table \ref{tab:classic_coupling} illustrates the coupling spectrum as defined implicitly in Structured Design, the Table \ref{tab:modern_coupling} presents explicitly the extended version of the same spectrum in~\cite{ Lethbridge-2005} with an emphasis on Object-Oriented Programming.

\begin{table}[h]
    \caption{Classic Coupling}
    \label{tab:classic_coupling}
    \centering
    \begin{tabular}{|p{1.2cm}|p{1.2cm}|p{1.2cm}|p{1.2cm}|p{1.2cm}|}
        \hline
        \multicolumn{5}{|c|}{\textbf{Type of Connection}} \\ \hline
        \multicolumn{1}{|c|}{Pathologically} & \multicolumn{2}{c|}{Normally} & \multicolumn{2}{c|}{Minimally} \\ \hline
        \multicolumn{5}{|c|}{\textbf{Interface Complexity}} \\ \hline
        \multicolumn{2}{|c|}{High} & \multicolumn{3}{c|}{Low} \\ \hline
        \multicolumn{5}{|c|}{\textbf{Binding Time}} \\ \hline
        \multicolumn{2}{|c|}{Early} & \multicolumn{3}{c|}{Late} \\ \hline
        \multicolumn{5}{|c|}{\textbf{Coupling}} \\ \hline
        Hybrid & Common & Content & Control & Data \\ \hline
        \multicolumn{2}{|c|}{Undesirable} & \multicolumn{3}{c|}{Desirable} \\ \hline
    \end{tabular}
\end{table}

We refer to the coupling spectrum as being defined implicitly because, unlike cohesion, which is explicitly detailed as a spectrum of similar types with varying degrees of desirability, coupling (intermodular relatedness) has been examined from multiple perspectives, such as \textit{connection type}, \textit{interface complexity}, \textit{binding time}.

This spectrum of coupling (loose-tight) has since been widely accepted by various others~\cite{ Lethbridge-2005, Page-Jones-1980, Gamma-1994, Pressman-2015, Newman-2015, Kent-2023}, although with slight variations (see Table~\ref{tab:modern_coupling}).

\begin{table}[h]
    \caption{Modern Coupling}
    \label{tab:modern_coupling}
    \centering
    \begin{tabular}{|p{1.9cm}|p{6.0cm}|c|}\hline
        \textbf{Type}            & \textbf{Description}   \\
        \hline
        Content  (most tight)   & A component is trying to spoof the internal data of another component. \\
        \hline
        Common                   & The use of global variables.  \\
        \hline
        Control                  & One procedure directly controls another using a flag. \\
        \hline
        Stamp                    & One of the argument types of a method is one of your application classes. \\
        \hline
        Data                     & The use of method arguments that are simple data. \\
        \hline
        Routine call             & A routine calling another.                  \\
        \hline
        Type use                 & The use of a globally defined data type.  \\
        \hline
        Inclusion/import         & Including a file or importing a package. \\
        \hline
        External (less tight - loose)    & A dependency exists on elements outside the scope of the system, such as the operating system, shared libraries, or hardware.  \\
        \hline
    \end{tabular}
\end{table}

More recent definitions of coupling do not differ much from those of other studies during 1970. For example, Grady Booch defines coupling in object-oriented design as the measure of the strength of association established by a connection from one module to another~\cite{Booch-2007}. 

Larman~\cite{Larman-2005} and Lethbridge~\cite{ Lethbridge-2005} mention that coupling occurs when there are interdependencies between one module and another". They provide three indications for interdependency, the \textit{rippling change} from the module that is under modification to all interdependent modules, the \textit{ difficulty to understand} the functionality of the coupled module and a higher effort in \textit{reuseability} due to essential physical dependencies. Table \ref{tab:modern_coupling} depicts the coupling spectrum according to
Lethbridge~\cite{ Lethbridge-2005}. 

Although we acknowledge that the rippling change may demonstrate the presence of coupling, we strongly disagree with the second indication because the term "understanding" is ambiguous and context dependent and also the presence of physical dependencies cannot necessarily conclude the presence of coupling~\cite{Suh-2001}. To further our discussion, we will dive into this topic in greater detail as we progress in this study. However, it is essential first to highlight the coupling issues found in other studies that have influenced the main argument of this research. 

DeMarco~\cite{DeMarco-1979} in his book states that: "There is no way to make modules in a structure absolutely independent of one another, but it is possible to come up with a structure that has so little coupling that you can usually modify one module without disrupting others." Myers~\cite{Myers-1975} also presented a similar argument, but more precisely argued that designing software systems that are fully independent of each other is unreachable. 

We contend that although the above-mentioned statements hold some truth, they reflect a misleading understanding of modularity and decoupling, which we intend to clarify here. Let us continue with DeMarco~\cite{DeMarco-1979}. Even though he attempted to soften the tone by using imprecise terms such as 'little' or 'disruption', it still accommodates a perfect paradox. Let us take a close look at what he says by focusing on these two terms "disruption" and "little coupling".

Based on fundamental principles of physics, disruption among two coupled elements could mean vibration, resonance, interference, thermal runaway, mechanical failure, etc. It is not clear what DeMacro means by disruption; however, from his book we could conclude that perhaps he means rippling change.

In order to proceed, we need to define the next term: "little". Similarly to the previous term, it also fails to provide significant insight for a software engineer to impact the design process. However, as derived from Section 25.2.3 (Coupling)~\cite{DeMarco-1979} in DeMarco's book, the quantity and nature of data transferred between modules are indicators of coupling strength, and the term "little" could mean minimization of data and perhaps some enhancements on the type of data. 

The paradox lies in the fact that as long as there is some form of coupling between elements (modules), regardless of how weak (loose) or strong (tight) it is, there remains a significant potential for disruption (rippling change). Additionally, the degree of coupling (the amount of data) does not have a linear relationship with the degree of disturbance. A small amount of data, shared among modules, could lead to large modifications, while large amounts of data can absorb small modifications without much disruption. It is the nature of data, along with its probability and frequency of being modified is the matter.

\subsection{Cohesion}
Britannica defines “Cohesion” as an attractive intermolecular force between molecules of the same kind or phase. Meriam Webster defines "cohesion" as the act or state of tightly binding. 

In 1968, during NATO conference, D.T. Ross~\cite{NATO-1968} used the term cohesive to describe part of his idea about "plex" that we covered earlier, exactly in the same meaning that later on others (e.g., Constantine, Meyer) used it. The following extracted text from his argument represents how he used the term cohesive.

During his argument about presenting "plex" idea in banking scenario, he states that: "Each integrated package is a sublanguage, processed systematically by performing lexical processing of the input string, performing syntactic and semantic parsing, building a model of the implications of the information, and acting on the information. This creates an idealized plex for each phase, mechanized and interlocked to form one cohesive and comprehensive language and system for banking problems."

This explains why we previously claimed that the concept of cohesion should actually be attributed to D.T. Ross instead of L.L. Constantine.

In 1974, Stevens, Myers, and Constantine published the paper titled "Structured Design"~\cite{Constantine-1974}. Subsequently, in 1978, Yourdon and Constantine used the same title for their book~\cite{Yourdon-1978}. In this publication, they described "cohesion" as a technique to enhance the relationships among elements (data, structure, and algorithm) within the same module. 

Then they introduced six distinct types of cohesion: coincidental, logical, temporal, communicational, sequential, and functional. By the time the "Structured Design" book was published in 1978, Yourdon and Constantine had introduced an additional type, called procedural. These types are presented in Table~\ref{tab:classic_cohesion}.

The authors of 'Structured Design' define "cohesion" as intermodular connections and the amount of intermodular coupling within a single module. They refuse to employ “intermodular relatedness” because they find this term clumsy. Therefore, they borrow the term "cohesion" from sociology~\cite{Yourdon-1978} to better formulate "intermodular relatedness (coupling)".

\begin{table}[h]
    \caption{Classic Cohesion~\cite{Yourdon-1978}}
    \label{tab:classic_cohesion}
    \centering
    \begin{tabular}{|p{2.4cm}|p{5.6cm}|c|}\hline
        \textbf{Type}            & \textbf{Description} \\
        \hline
        Functional (Spectrum: Highest)  & Module contains elements that all contribute to the execution of one and only one problem. \\
        \hline
        Sequential                   & The one whose elements are involved in activities such that the output data of one activity serves as input data to the next activity. \\
        \hline
        Communicational                  & The one whose elements correspond to activities that use the same input or output data.\\
        \hline
        Procedural                    & Result of flowchart based modularization, sequence unrelated activities in one module.\\
        \hline
        Temporal                     & Same as logical, except that there is time-relatedness as well. \\
        \hline
        Logical             & Implies some logical relationship among the elements of a module. \\
        \hline
        Coincidental (Spectrum: Lowest)  & Occurs when there are no meaningful relationships among the elements. \\
        \hline
    \end{tabular}
\end{table}

Yourdon and Constantine provided a very peculiar definition of functional cohesion: Anything that is not sequential, communicational, procedural, temporal, logical, or coincidental is functional cohesion. Page-Jones articulates a deep definition of functional cohesion, describing it as a condition where every element within a module works together to solve a single problem (note the idea behind the "plex").~\cite{Page-Jones-1980}\footnote{Definition of Page-Jones from functional cohesion closely relates to Unix-philosophy discussed by McIlroy~\cite{McIlroy-1978} in 1978.}.

It is fascinating to observe how the initial theorists on "cohesion" were unsure about the terminology to characterize its essence:
\begin{itemize}
    \item In 1974, Stevens, Myers, and Constantine used the term "binding" to define different degrees of cohesion~\cite{Constantine-1974}.
    \item In 1975, Myers used the term "strength" and eliminated the term "cohesion"~\cite{Myers-1975}.
    \item In 1978, Yourdon and Constantine used the term "association" to define different degrees of "cohesion"~\cite{Yourdon-1978}.
    \item In 1979, 1980 and 1988, DeMarco and Page-Jones together used the terms "strength" and "association" to talk about "cohesion" and its degrees~\cite{DeMarco-1979,Page-Jones-1980,Page-Jones-1988}. 
\end{itemize}

We argue that in principle, "cohesion" is in fact a intra-modular coupling that the designer of system deliberately considers in his design to maximize integrity, predictability, maintainability, and interchangeability of modules and avoid creation of inter-modular coupling. In fact, by understanding the concerns of the system in depth and considering them during modularization, we eventually reach high cohesion (functional cohesion) if the cost and system constraints justify its necessity.

\section{Low-Coupling and High-Cohesion}
\label{Coupling_and_Cohesion}
Numerous renowned authors~\cite{DeMarco-1979, Page-Jones-1980, Fowler-2001, Evans-2003, Larman-2005, Lethbridge-2005, Pressman-2015, Newman-2015, Kent-2023} accept the "high cohesion and low coupling" as a guiding principle to build a good software. The only noticeable change among their works has been in \textit{terminology} (e.g., module, component, service, object, class). 

For example, "Domain-Driven Design" by Eric Evans, "Building Microservices" by Sam Newmann~\cite{Newman-2015}, "Tidy First" by Kent Beck~\cite{Kent-2023} and other scholars like Roger S. Pressman~\cite{Pressman-2015}, C. Larman~\cite{Larman-2005}, and T. Lethbridge and R. Laganiere employ this principle extensively at their work without questioning it in the slightest. 

However, there are also others who question it and argue that this principle does not result in an efficient design~\cite{Brito-2001, AlDallal-2011, Candela-2016, Paixao-2018} or notably deviate from this principle. 

For example, Meyer~\cite{Meyer-1997b} and Booch~\cite{Booch-2007} argue that the original definitions of coupling and cohesion by structured design are influenced by the traditional concept of a subroutine, thus narrowing the discussion. 

Even Yourdon and Constantine in their summer of chapter 7~\cite{Yourdon-1978} state:
"High cohesion is not "good", nor is coincidental cohesion "evil", module cohesion is associated with effective modularity, and it has certain predictable effects on transparency, programmability, ease of debugging, ease of maintenance, and ease of modification."

Martin~\cite{R.C.Martin-2017} in "Clean Architecture" avoids employing this principle in its original version ("high cohesion and low coupling") and instead concentrates on component cohesion, coupling, and decoupling as design methods to improve software quality in terms of changeability, testability, and deployability. He advocates for the application of these techniques and principles at the appropriate moments and locations, carefully balancing any potential conflicts that may arise, i.e. he neither says go always for high cohesion nor low coupling rather than balance them based on needs and concerns of project and system.

Another interesting finding is that a detailed review of GoF shows that the term "Cohesion" is never mentioned in the entire book~\cite{Gamma-1994}, with all patterns governed by the coupling spectrum, be it loose or tight. 

We believe that the term cohesion does not appear in the GoF due to an accident or editorial mistake, but for a valid reason: in software, we are always concerned with the spectrum of coupling (various types of coupling). Whether one considers it useful to view the coupling spectrum or simply classify it as either coupled or decoupled, the argument is that whenever a self-contained and decoupled module can be created, the design is ideal and requires no further modifications. Otherwise, it becomes crucial to employ specific design patterns aimed at either minimizing coupling or reducing its impact (promoting loose coupling).

\section{The Chapter 5}
\label{Fundamental Theory}
It is not a slip of the pen that this section is named Chapter 5. This fundamental theorem is often overseen by others, leading them to blindly argue on decomposition without considering its cost. In 1978, Yourdon and Constantine formulated the Fundamental Theorem of Software Engineering~\cite{Yourdon-1978} based on G. Miller's work from 1956~\cite{Miller-1956}. 

Miller posited in his paper that whether it is a mere Pythagorean coincidence or whether there is a deeper significance to number 7, the immediate recirculating memory capacity for problem solving with multiple elements is constrained to around 7±2 entities. As articulated in~\cite{Yourdon-1978}, regardless of whether these elements consist of state variables, subroutine calls, or other factors, errors escalate sharply and nonlinearly once this 5-9 element limit is exceeded. This represents a fundamental characteristic of human information processing, which has led to the practice of segmenting, factoring, and decomposing problems into subproblems.

In this chapter based on human input processing capability, the Fundamental Theorem of Software Engineering has been crafted, and surprisingly this part has been often overseen by other scholars and practitioners mainly by~\cite{Fowler-2001,Evans-2003,Larman-2005, Lethbridge-2005, Kent-2023}:

\begin{equation}
    C(P) > c\left(\frac{1}{2}p\right) + c\left(\frac{1}{2}p\right)
    \label{eq:equation1}
\end{equation}

Essentially, the theorem indicates that addressing problem \textit{P} is more costly compared to splitting it into two independent sub-problems, if and only if each sub-problem could be addressed independently. Otherwise, introducing interactions between subproblems, which could potentially increase overall cost to higher cost than the original problem, may be seen to be unavoidable (see Eq. \ref{eq:equation2}):

\begin{equation}
    c(p' + I_1 \times p'') + c(p'' + I_2 \times p') > c\left(\frac{1}{2}p + \frac{1}{2}p\right)
    \label{eq:equation2}
\end{equation}

Let`s assume that each subproblem is a component and each interface in between an interrelatedness (amount of information that each component should share with the related component). The problem with introducing new interfaces is that it could easily result in rippling changes from one component to another component(s), and/or it could increase the information speed and velocity among components, i.e. the cost of introducing a change may increase as the change propagates among all other related components, it could increase information velocity and speed, and could result in higher information exchanging~\cite{Meyer-1997b}. 

\section{Harm of Cohesion}
\label{HarmOfGood}
Although aiming for high cohesion is not necessarily harmful, it has potential to lead to a harmful state of being if it is done without considering its cost. Throughout this study, we used separate occasions to support our argument. Let us present you with an example one more time why we say that striving for high cohesion or avoiding low cohesion could be harmful.

Consider having a set of operations that are independent in the sense that each can perform without the presence of others, but there is an order among them that has to be followed for a business reason (i.e. due to non-functional requirements). Gathering all these operations into one module will result in procedural cohesion~\cite{Page-Jones-1980} which is not a desired level of cohesion, but separating each one into one independent module (e.g., class, service, microservice) will bring functional cohesion at the cost of introducing an interlocking mechanism (orchestration), which could significantly contribute to the complication and cost of the system. A similar conflict could be observed when one tends to employ REP (Release-Reuse Equivalence Principle).

Consider a more practical example from a recent industrial case of Amazon Prime Video, where they transitioned back from a distributed microservice architecture to a monolithic application, achieving a 90\% cost savings~\cite{Amazon-2024}.

In another example, let us refer to a part of the book recently authored by Kent Beck: "Tidy First?"~\cite{Kent-2023}. In this book he states that cohesion has two implications:
First, "coupled elements should be subelements of the same containing element", and the "second implication of cohesion is that elements that are not coupled should go elsewhere". Although the first implication presents the same idea as expressed by Constantine and others~\cite{Constantine-1967,Yourdon-1978, DeMarco-1979,Page-Jones-1980} the second implication actually violates what was discussed in Chapter 5 of Structured Design as the Fundamental Theory of Software Engineering (see Eq~\ref{eq:equation1} and Eq.~\ref{eq:equation2}). The term "elsewhere" in the second implication eventually could result in the introduction of a new element resulting in a higher cost due to intermodular communication, extra effort in deployment, and dependency resolution effort.\footnote{For example think of opting for Reuse/Release Equivalence Principle (REP) while trying to achieve high cohesion.}

\section{Redefining Good Software}
In this section, our objective is to encapsulate all our results as explained in the following corollaries and, subsequently, propose a redefined description of good software based on these corollaries.

\subsubsection{Corollary 1}
Low cohesion does not necessarily imply that we should split the module (for instance, one with coincidental cohesion) into separate modules. This is because they might still need to communicate, and introducing interfaces between them could replace in-memory communication with more costly alternatives such as socket. We should divide the module only if we can establish two completely independent (decoupled) modules or prove that the cost of additional interfaces will not exceed the original cost (Fundamental Theory of Software Engineering~\cite{Yourdon-1978}).

\subsubsection{Corollary 2}
Based on the literature provided, we can conclude that in fact the definition of good software (high cohesion and low coupling) accepts the essence of coupling in software systems but aims to reduce it. While this is partially true, the determination of the adequacy of low coupling is highly subjective and cannot be effectively guided by the coupling and cohesion spectrum which is a very algebraic approach.

\subsubsection{Corollary 3}
We argue that while reducing coupling is often a good strategy to improve evolutionary aspects of software, there are also occasions that we may aim for higher coupling (e.g., increase of predictability or stability as suggested by SDP~\cite{R.C.Martin-2000}). Additionally, cohesion itself represents a different form of coupling that happens in different locations and forms a desirable coupling that could result in several improvements, such as information hiding. 

\subsubsection{Corollary 4}
Although modularization in software systems begins with decomposition, the designer can eventually utilize composition and oscillate between different levels of granularity, depending on the concerns guiding the design of the system with the goal of reaching an acceptable balance among the varying concerns.
 
\subsubsection{Corollary 5}
With the computational capabilities that we have today, the challenges mentioned in 1950~\cite{Tompkins-1950} would not be deemed complex problems. Therefore, if we define complexity solely within the physical domain, such as Tompkins~\cite{Tompkins-1950} and Dijkstra~\cite{Dijkstra-1965} and tend to use modularization to tackle it, there will be varying interpretations of complexity~\cite{Suh-2005} that lead to different modularization depending on who is doing the modularization. Therefore, it is important to understand that, first, divide-and-rule is not the topic of modularization, second modularization should not be based on defined complexity in the physical domain, as such definition of complexity will have different interpretations depending on whom we ask~\cite{Suh-2001}.

\subsubsection{Corollary 6}
It is debatable whether software systems should be viewed as evolving entities or merely as a product of natural selection. However, it is crucial to remember that the fundamental characteristics of software systems should not be compromised or diminished during any activities such as modification or reconstruction. In other words, software systems must undergo inherited modifications with respect to their primary concerns. Thus, if we consider modularization as a strategy to extend the lifespan of software systems, it is essential to align it with the system's purpose. For example, if changeability is not the primary concern for a software system compared to predictability, then the system designer should focus on modularization aimed at enhancing reliability rather than changeability.   

Considering the above corollaries, we define good software as one that is modularized in the direction of its main concerns.

\section{Conclusion}
In conclusion, the exploration of modularization, coupling, and cohesion in software engineering reveals several critical insights. First, low cohesion within a module does not automatically warrant its division into separate modules unless true independence can be achieved, as otherwise, the costs of inter-module communication could outweigh the benefits. 

Additionally, the conventional wisdom that good software design is defined by high cohesion and low coupling is found to be subjective and potentially misleading. The simplistic acceptance of coupling as an inevitable condition should be revisited, recognizing that certain forms of coupling, such as those guided by the Stable Dependencies Principle, can be beneficial. Moreover, cohesion itself represents a valuable form of coupling that can lead to significant improvements in software design, such as enhanced information hiding. 

Furthermore, modularization is not merely a process of decomposition; it requires a dynamic approach where designers oscillate between different levels of granularity to balance various design concerns effectively. 

Finally, the long-term sustainability of software systems requires that any modifications (such as modularization) respect the core concerns of the system. Modularization should therefore align with the purpose of the system, for example, ensuring that characteristics such as predictability or reliability are prioritized over changeability or reusability in systems whose main purpose is to provide predictable and reliable functionality such as the controller a luggage handling system in an airport. 

An airport luggage handling system does not necessitate the same level of changeability as other commercial software systems (e.g., social networks, accounting, taxing); however, it is crucial for such a system to deliver reliable and predictable functionality.

Further research could be conducted to gain a clearer understanding of the concept of evolutionary software and to determine if it is appropriate to view software systems as evolving entities. 

\section*{Acknowledgments}
I am deeply grateful to Mr. Robert C. Martin for his valuable feedback on the paper, which helped me fine-tune several arguments. Furthermore, I wish to express my thanks to Prof. Dr. Mario de Sousa for posing critical questions and prompting me to reconsider my thoughts.

\bibliography{references}
\bibliographystyle{IEEEtran}

\newpage


\vfill

\end{document}